\documentclass[conference, 9pt]{IEEEtran}
\IEEEoverridecommandlockouts
\usepackage{cite}
\usepackage{amsmath,amssymb,amsfonts}
\usepackage{algorithmic}
\usepackage{graphicx}
\usepackage{textcomp}
\usepackage{xcolor}
\def\BibTeX{{\rm B\kern-.05em{\sc i\kern-.025em b}\kern-.08em
    T\kern-.1667em\lower.7ex\hbox{E}\kern-.125emX}}

\usepackage{subfigure}

\begin{document}

\title{Machine Learning aided Computer Architecture Design for CNN Inferencing Systems\\
\thanks{This work was partly supported by the Data Science Center of the University of Bremen (DSC@UB)}
}

\author{\IEEEauthorblockN{Christopher A. Metz}
\IEEEauthorblockA{
\textit{Supervisor: Rolf Drechsler}\\
\textit{Institute of Computer Science, University of Bremen}\\
Bremen, Germany \\
cmetz@uni-bremen.de} 
}

\makeatletter
\def\ps@IEEEtitlepagestyle{%
  \def\@oddfoot{\mycopyrightnotice}%
  \def\@oddhead{\hbox{}\@IEEEheaderstyle\leftmark\hfil\thepage}\relax
  \def\@evenhead{\@IEEEheaderstyle\thepage\hfil\leftmark\hbox{}}\relax
  \def\@evenfoot{}%
}
\def\mycopyrightnotice{%
  \begin{minipage}{\textwidth}
  \scriptsize
  Copyright is held by the author/owner(s).\\
  \end{minipage}
}
\makeatother

\maketitle

\begin{abstract}
Efficient and timely calculations of Machine Learning (ML) algorithms are essential for emerging technologies like autonomous driving, the Internet of Things (IoT), and edge computing. One of the primary ML algorithms used in such systems is Convolutional Neural Networks (CNNs), which demand high computational resources. This requirement has led to the use of ML accelerators like GPGPUs to meet design constraints. However, selecting the most suitable accelerator involves Design Space Exploration (DSE), a process that is usually time-consuming and requires significant manual effort. \\
Our work presents approaches to expedite the DSE process by identifying the most appropriate GPGPU for CNN inferencing systems. We have developed a quick and precise technique for forecasting the power and performance of CNNs during inference, with a MAPE of 5.03\% and 5.94\%, respectively. Our approach empowers computer architects to estimate power and performance in the early stages of development, reducing the necessity for numerous prototypes. This saves time and money while also improving the time-to-market period. 
\end{abstract}


\section{Introduction}
Advancements in Machine Learning (ML) and Artificial Intelligence (AI) have yielded impressive results. To ensure fast and efficient calculations, AI accelerators are increasingly being utilized. The latest developments include GPGPUs designed specifically for ML training and inferencing, which can consume up to 700 watts per GPGPU. As most High-Performance Computing (HPC) systems come equipped with multiple GPUs per machine, the power consumption of ML and AI systems presents new challenges \cite{Metz.2021.CODESISSS, Metz.2023.PAISE,Metz.2021.SLOHA,Metz.2022.DDECS,Metz.2022.MLCAD }.

An example of extreme power consumption in HPC can be seen in the Summit, a supercomputer with 27,648 NVIDIA Volta GPUs that consume 13 million watts \cite{Foertter2018}. However, by implementing power savings of 5\%, significant cost savings of up to 1 million dollars can be achieved \cite{Guerreiro2019}. On the other hand, smaller Internet of Things (IoT) devices can also experience increased power consumption due to ML inferencing. For instance, executing object recognition on an Nvidia Jetson TX1 can consume 7 watts, but offloading the same task to the cloud reduces power consumption to 2 watts. Therefore, offloading can be a promising strategy in ML-enabled IoT applications with limited battery resources. In certain situations, accomplishing tasks locally or offloading them may require distinct strategies. Additionally, the feasibility of offloading ML workloads depends on available bandwidth. Therefore, local execution may be necessary when offloading is not a viable option. Due to the large design space, these challenges make it difficult for computer architects to design appropriate ML inferencing systems. 
 
There are different ways to explore design space for ML inferencing computer architecture, but two main approaches stand out: 1)~simulation and 2)~ML-based predictors. However, both have their drawbacks. For example, simulators like GPGPU-Sim or GPU-ocelot run GPU applications on CPUs for simulation, which leads to significantly slower simulations than on real devices due to CPUs not having the same high parallelization ability as GPUs. ML-based predictors aim to provide fast and accurate estimations, but most require specific configuration and profiling of the application on a real GPGPU first to collect performance counters. Since performance counters are not standardized across all Nvidia GPUs, it's possible that the required counter is unavailable or is collected differently than in the original approach, making it impossible to apply the approach or result in inaccurate results\cite{Metz.2023.FDL}. However, none of these approaches consider the option to offload the ML workloads to cloud or edge systems.

This work addresses the limitations of simulation and current ML-based predictors for predicting the power and performance of ML inferencing on GPGPUs. Our study presents recent approaches to overcome these obstacles. Our contributions mainly include the following:
\begin{itemize}
    \item Our ML-based power and performance estimation is both fast and accurate \cite{Metz.2023.PAISE, Metz.2021.CODESISSS, Metz.2022.MLCAD, Metz.2022.DDECS, Metz.2021.SLOHA}.
    \item We developed a hybrid PTX Analyzer that can collect runtime-dependent features for power and performance estimation without executing on real GPUs. This solution should also overcome the slow execution time of simulators \cite{Metz.2023.FDL}.    
\end{itemize}

\section{Power and Performance Estimation}

\begin{figure}[t]
	\centering
	\includegraphics[width=.8\columnwidth]{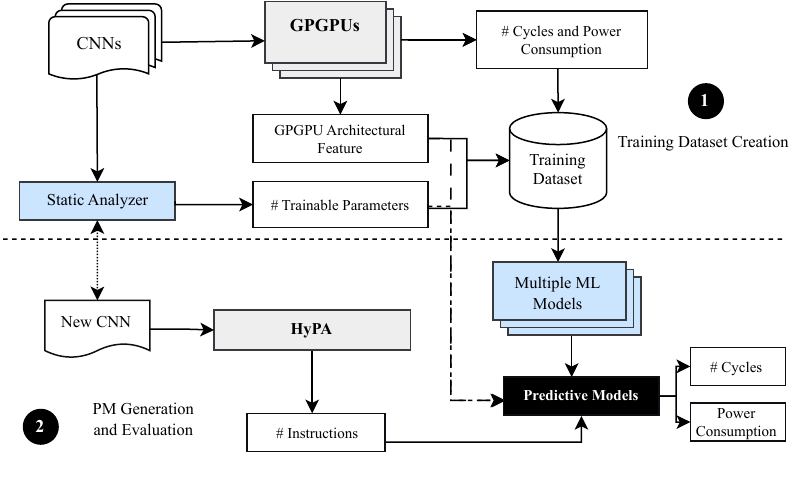}
	\caption{Methodology for estimating performance and power has been adapted from \cite{Metz.2023.PAISE}.}
	\label{fig:Method}
 \vspace{-.5cm}
\end{figure}

\begin{figure*}[htb]
\centering
	\subfigure[Efficientnetb0]{\includegraphics[width=.27\textwidth]{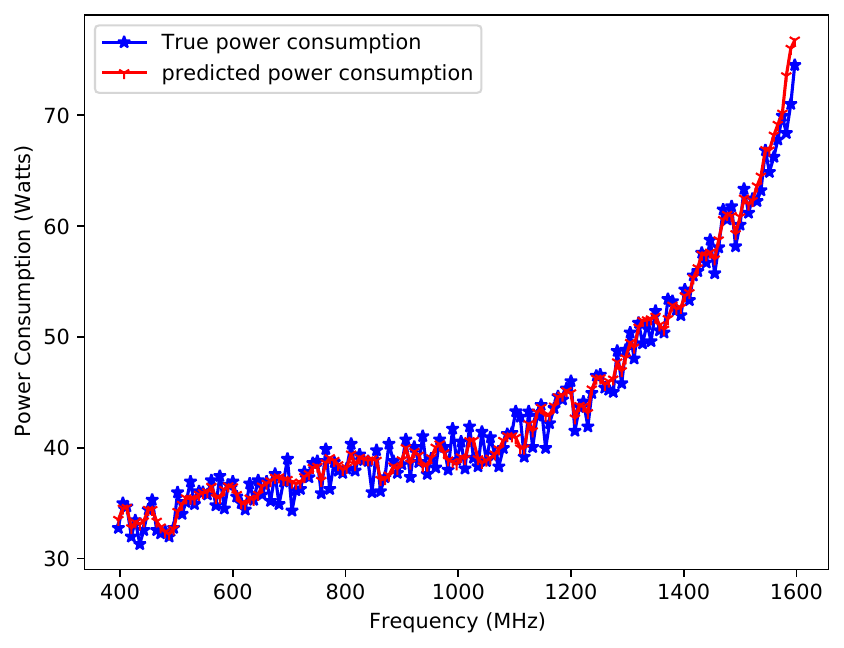}}
	\subfigure[Efficientnetb1]{\includegraphics[width=.27\textwidth]{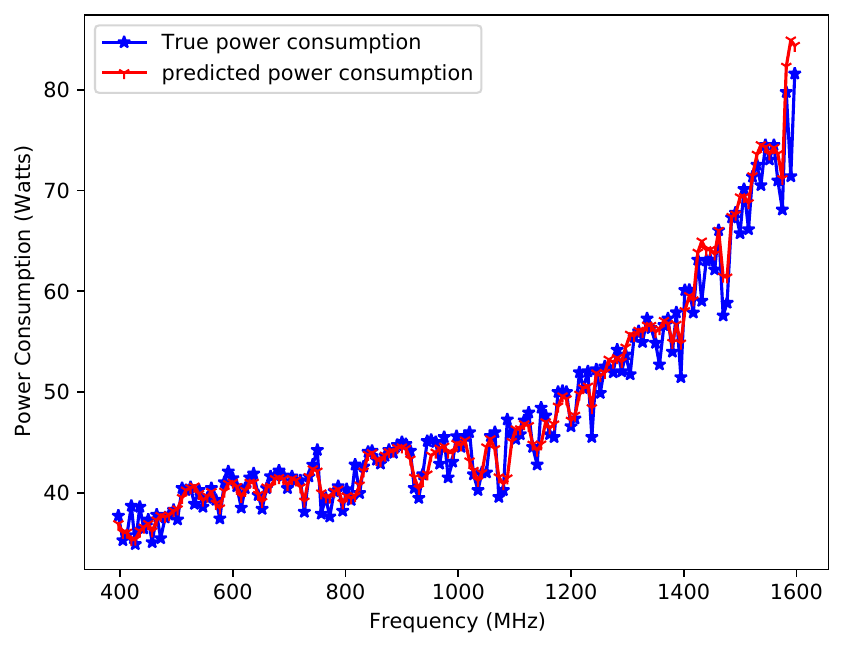}}
	\subfigure[Efficientnetb2]{\includegraphics[width=.27\textwidth]{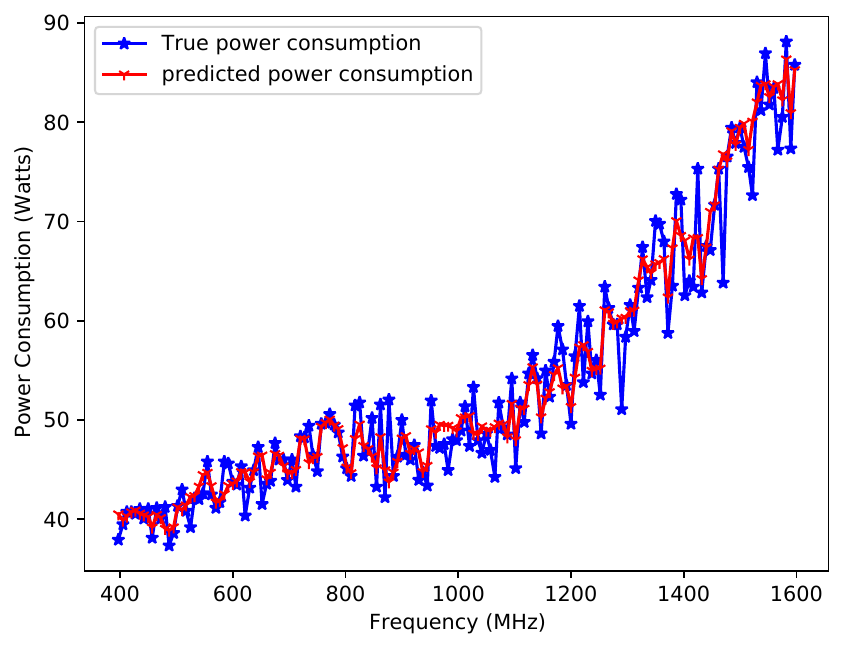}}
	\caption{Comparison of predicted and real power consumption for three CNNs with different frequencies between 397MHz and 1590MHz on the Nvidia V100S GPGPU \cite{Metz.2022.MLCAD}}
	\label{fig:predictiveResults}	
 \vspace{-.5cm}
\end{figure*}

Figure \ref{fig:Method} briefly overview our process for estimating Power and Performance when developing ML models. To ensure accurate results, we train multiple machine learning models (e.g., K-Nearest Neighbor, Decision Tree, Random Forest Tree) for each specific task (i.e., power or performance prediction), which helps improve each model's accuracy. As predictions must be made during the early design stages, we focus on not runtime-dependent features. This includes utilizing hardware specifications such as the size and factor of the GPGPU, the number of cores, the frequency, and the available memory. Additionally, we consider features that describe the ML application (e.g., neural networks) that consist of varying layers and neurons. This approach allows us to develop ML models that are both effective and efficient \cite{Metz.2021.CODESISSS, Metz.2021.SLOHA, Metz.2022.DDECS,Metz.2022.MLCAD,Metz.2023.PAISE}. 

Additionally, we have created a new tool called \emph{Hybrid PTX Analyzer}~(HyPA) to account for the intricacies of the compiled ML Model. This tool lets us determine the exact number of executed instructions in the PTX without running the code on physical devices. To achieve this, we simulate critical code sections such as loops or if-statements to construct an accurate control flow graph that encompasses all necessary instructions. Thus, we can also consider runtime-dependent features without executing GPU applications on actual devices \cite{Metz.2023.FDL}.  

\section{Experimetal Results.}
In the following, we present power and performance estimation results of ML model inferencing on GPGPUs based on \cite{Metz.2021.CODESISSS, Metz.2022.DDECS,Metz.2021.SLOHA,Metz.2022.MLCAD,Metz.2023.PAISE}.
The power prediction for various frequencies of the Nvidia V100S is depicted in fig. \ref{fig:predictiveResults} \cite{Metz.2022.MLCAD}. In our studies, the Random Forest Trees achieve a \emph{Mean Absolute Percentage Error}~(MAPE) of 5.03\% and a $R^2$-Score of 0.9561 for the power prediction for different CNNs at different frequencies. This methodology can also be applied to other GPUs, enabling the creation of predictive models for each GPU that can forecast power usage for different Neural Networks at varying frequencies.

\begin{figure}[htb!]
	\centering
	\includegraphics[width=.8\columnwidth]{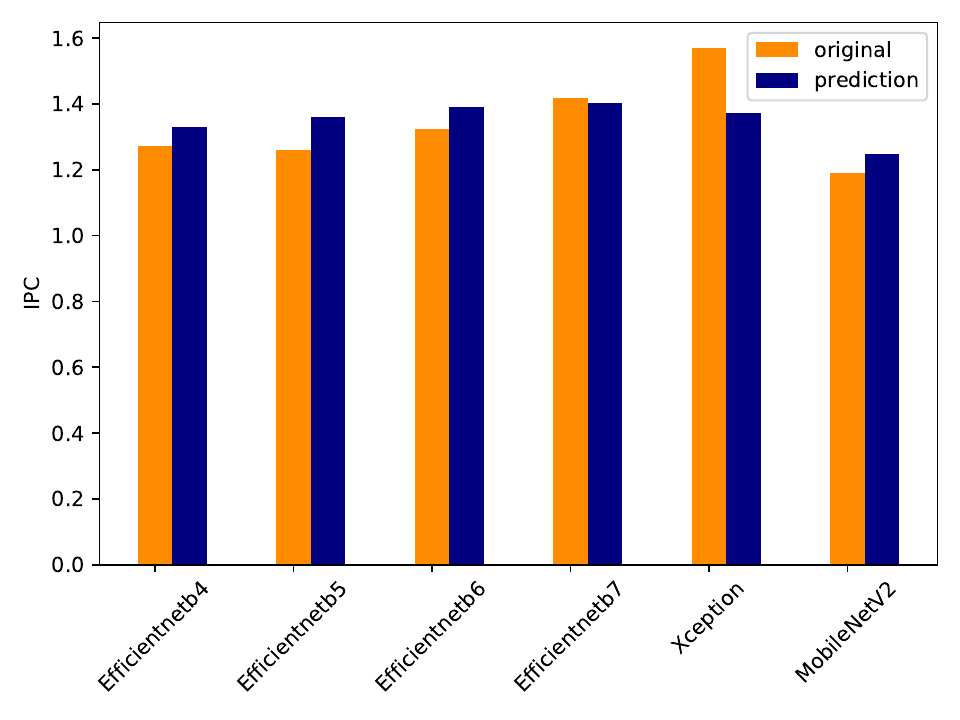}
	\caption{Prediction results for number of cycles \cite{Metz.2023.PAISE}.}
	\label{fig:PerformancePrediction}
\end{figure}

The performance prediction (i.e., number of cycles) for different Neural Networks is illustrated in fig. \ref{fig:PerformancePrediction}. As the results demonstrate, the K-Nearest Neighbors Algorithm achieved a MAPE of 5.94\% \cite{Metz.2023.PAISE}.

The results show that our methodology allows the generation of fast and accurate predictive models for estimating power and performance. This is beneficial for computer architects in navigating the design space and identifying the optimal GPGPU. 

\section{Conclusion and Future Work}
In our upcoming projects, we aim to incorporate optimization techniques to search for the best GPGPU to enhance ML model inference while considering factors such as limited power supply and desired performance. Additionally, we intend to devise approaches to discern whether offloading would adhere to the constraints or if executing locally would be more advantageous. We have developed a REST API for offloading ML workloads and are currently studying the power and performance characteristics at various bandwidths and latencies. 

We plan to merge power and performance prediction for GPGPUs with the findings from our offloading analysis. This will help us identify the most suitable GPGPU for local execution. As a result, computer architects and system designers will be able to reduce the number of prototypes they need to build. 

\bibliographystyle{IEEEtran}

\bibliography{ref}

\begin{thebibliography}{1}
\providecommand{\url}[1]{#1}
\csname url@samestyle\endcsname
\providecommand{\newblock}{\relax}
\providecommand{\bibinfo}[2]{#2}
\providecommand{\BIBentrySTDinterwordspacing}{\spaceskip=0pt\relax}
\providecommand{\BIBentryALTinterwordstretchfactor}{4}
\providecommand{\BIBentryALTinterwordspacing}{\spaceskip=\fontdimen2\font plus
\BIBentryALTinterwordstretchfactor\fontdimen3\font minus
  \fontdimen4\font\relax}
\providecommand{\BIBforeignlanguage}[2]{{%
\expandafter\ifx\csname l@#1\endcsname\relax
\typeout{** WARNING: IEEEtran.bst: No hyphenation pattern has been}%
\typeout{** loaded for the language `#1'. Using the pattern for}%
\typeout{** the default language instead.}%
\else
\language=\csname l@#1\endcsname
\fi
#2}}
\providecommand{\BIBdecl}{\relax}
\BIBdecl

\bibitem{Metz.2021.CODESISSS}
C.~A. Metz, M.~Goli, and R.~Drechsler, ``{Early Power Estimation of CUDA-Based
  CNNs on GPGPUs: Work-in-Progress},'' in \emph{Proceedings of the 2021
  International Conference on Hardware/Software Codesign and System Synthesis},
  ser. CODES/ISSS '21.\hskip 1em plus 0.5em minus 0.4em\relax New York, NY,
  USA: Association for Computing Machinery, 2021, p. 29–30.

\bibitem{Metz.2023.PAISE}
------, ``{Fast and Accurate: Machine Learning Techniques for Performance
  Estimation of CNNs for GPGPUs},'' in \emph{2023 IEEE International Parallel
  and Distributed Processing Symposium Workshops (IPDPSW)}, 2023, pp. X--Y.

\bibitem{Metz.2021.SLOHA}
------, ``{Pick the Right Edge Device: Towards Power and Performance Estimation
  of CUDA-based CNNs on GPGPUs},'' \emph{CoRR}, vol. abs/2102.02645, 2021.

\bibitem{Metz.2022.DDECS}
------, ``{ML-based Power Estimation of Convolutional Neural Networks on
  GPGPUs},'' in \emph{2022 25th International Symposium on Design and
  Diagnostics of Electronic Circuits and Systems (DDECS)}, 2022, pp. 166--171.

\bibitem{Metz.2022.MLCAD}
------, ``{Towards Neural Hardware Search: Power Estimation of CNNs for GPGPUs
  with Dynamic Frequency Scaling},'' in \emph{Proceedings of the 2022 ACM/IEEE
  Workshop on Machine Learning for CAD}, ser. MLCAD '22.\hskip 1em plus 0.5em
  minus 0.4em\relax New York, NY, USA: Association for Computing Machinery,
  2022, p. 103–109.

\bibitem{Foertter2018}
F.~Foertter, ``{Summit GPU Supercomputer Enables Smarter Science},''
  \url{https://developer.nvidia.com/blog/summit-gpu-supercomputer-enables-smarter-science/},
  2018, accessed: 22.03.2022.

\bibitem{Guerreiro2019}
J.~Guerreiro, A.~Ilic, N.~Roma, and P.~Tomás, ``{GPU Static Modeling Using PTX
  and Deep Structured Learning},'' \emph{IEEE Access}, vol.~7, pp.
  159\,150--159\,161, 2019.

\bibitem{Metz.2023.FDL}
C.~A. Metz, C.~Plump, B.~J. Berger, and R.~Drechsler, ``{HyBrid PTX Analysis
  for GPU accelerated CNNs inferencing aiding Computer Architecture Design},''
  in \emph{2023 Forum on Specification \& Design Languages (FDL)}, 2023,
  accepted for publication.

\end{thebibliography}
\end{document}